\documentclass[a4paper,12pt,preprint,aps,floatfix]{revtex4}
\usepackage[utf8]{inputenc}
\usepackage{longtable}
\usepackage{natbib}
\usepackage{graphicx}     
\usepackage{lscape}
\usepackage{color}
\usepackage{amsmath}
\usepackage{array}
\usepackage{pdflscape}

\newcommand{\cm}{\text{cm}$^{-1}$}

\newcommand{\CS}{$^{12}$C$^{32}$S$_2$}

\newcommand{\nbREF}{21}
\newcommand{\nbEL}{4279}
\newcommand{\nbTR}{8775}

\newcommand{\nbUTR}{7430}

\begin{document}
%\hspace{3cm}

\title{Empirical rovibrational energy levels for carbon disulfide}

\author{Tanvi Sattiraju and Jonathan Tennyson$^{*}$} 
\affiliation{Department of Physics and Astronomy, University College London, Gower Street,  London WC1E~6BT, UK.}

\email{j.tennyson@ucl.ac.uk}

\date{\today}

\label{firstpage}

\begin{abstract}%{We need some keywords}
An analysis of the measured rovibrational transitions is carried out for the \CS\ isotopologue of carbon disulfide. Data from 21 sources is extracted and validated using a consistent set of standard linear molecule quantum numbers. A corrected list of 8714 CS${_2}$ transitions forms the input to a Measured Active Rotational–Vibrational Energy Levels (MARVEL) procedure, generating \nbEL\ empirical rovibrational energy levels across 138 bands of \CS. Results are compared to the recent NASA Ames line list. While the agreement is generally good, issues
are identified with the energy levels of some states, notably those with
high values of the $v_2$ bending quantum number.
\end{abstract}

\maketitle

\section{Introduction}
Carbon disulfide, CS$_2$,  is a trace atmospheric species which largely comes naturally from  volcanoes and from combustion, mainly due to human activity. Since 2020, CS$_2$ has
been included in the HITRAN data base \cite{jt857}. CS$_2$ is flammable and its spectrum is the subject of combustion studies \cite{22PeNiCh.CS2}.
CS$_2$ has been detected in comets \cite{04JaScXu.CS2,16CaAlBa.CS2} and the atmosphere of Jupiter
following the impact of comet Shoemaker-Levy 9 \cite{95AtEdTr.CS2}. Recently,  CS$_2$ was detected   in the atmosphere of Venus
by the Venus Express mission \cite{23MaRoMi.CS2} and evidence for CS$_2$ in the sub-Neptune exoplanet TOI$-$270 d  was reported by Holmberg and Madhusudhan \cite{24HoMa.CS2} based on observations using the James Webb Space Telescope (JWST). As discussed in detail below, \CS, and indeed its various isotopologues, have been extensively
characterized in the laboratory using high resolution spectroscopy 
\cite{20KaGoKoMu,
12VaTsLeHe,
71SmOv,
00BlWaBrDu,
96PlMaHoDe,
73Maki,
79JoKaAn,
84BaBlWaCo,
85BlBaCaWa,
85LiJo,
92BlWaBlBr,
92WaMaBl,
99BlWaBrDu,
01BlWaBrDu,
04HoAnPiAl,
70SmOv,
74MaSa,
80JoKa,
88WeScMa,
86DaBlWaCo,      
99LiHu}. 
CS$_2$ has also been the subject of a recent
global effective Hamiltonian analysis \cite{22Tashkun} and a room-temperature, variational-nuclear-motion  line list
\cite{24HuGoTa.CS2}. 

In this paper we perform a comprehensive MARVEL (Measured Active Rotational–Vibrational Energy Levels) study of 
\CS. The MARVEL methodology has already been successfully used for OCS \cite{jt916} and is being actively
applied to parent CO$_2$ ($^{12}$C$^{16}$O$_2$) \cite{jt963} alongside its various isotopologues \cite{jt925,jt932,jt955}.
There are number of advantages of these studies, not least is the large increase in the number of transitions
who line frequencies are known to experimental accuracy at the end of the procedure \cite{jt828,jt923}.
In the case of OCS, the MARVEL study \cite{jt916} was used by the ExoMol project \cite{jt528,jt939} to construct a new line list for
hot OCS \cite{jt943} which has very recently been used to tentatively identify the signature of OCS in the atmosphere of hot Jupiter WASP-15b 
using the James Webb Space Telescope \cite{24KiAhCl.OCS}. The purpose of the current study is to provide the energy level input
needed for the construction of the first high-temperature line list for  CS$_2$ as part of the ExoMol project.

\section{Method}

\subsection{The MARVEL procedure}
The MARVEL algorithm \cite{jt412,13FuSzMa.marvel,jt750} is based on the theory of spectroscopic networks (SNs) \cite{11CsFuxx.marvel}. A molecular SN is described by a graph $G(E, f)$ with the nodes $E$ being the set of energy levels and the edges $f$ the set of transitions. On an input of the set $f$, MARVEL will generate the set $E$. The MARVEL procedure is carried out as follows:
\begin{enumerate}
    \item Experimentally measured transitions are extracted from various sources and compiled in a grand database. This includes the experimentally derived uncertainties and unique quantum number labels associated with each transition.
     \item Each transition from the cleaned database is given a tag detailing its source, and its numerical position within the source. 
    \item A spectroscopic network is established based on the input grand database.
    \item The sources are critically evaluated for inconsistencies and mislabelling in transitions collected from different sources. An attempt is made to correctly reassign the quantum numbers of the transition if the error is obvious to spot. 
    \item Some data sources mention lines poorly determined by experiment. In the case of such labelled 'weak' or 'blended' lines, the uncertainty is systematically increased.
    \item Transitions that do not validate are retained in the compilation but given a negative frequency and not considered by MARVEL. Their possible re-inclusion is reviewed as the final step in the MARVEL process which can be done automatically using the revive facility in MARVEL4.
    \item MARVEL sets up a vector of transitions within the spectroscopic network and carries out a matrix inversion to obtain the vector of energy levels.
    \item Experimental uncertainties are incorporated into the calculation. The treatment of uncertainties in MARVEL has varied between the versions. The latest version, MARVEL4 \cite{jt908,jt915,jt916}, used here employs a bootstrap method for assigning uncertainties to the elements of the energy level vector.
\end{enumerate}

\subsection{Uncertainty - the bootstrap method}
Questions can be asked regarding the validity of experimentally prescribed transition line uncertainties, and how these might translate to uncertainties in the empirical energy levels. The reported accuracy of high resolution spectroscopic experiments, and the subsequent determination of the associated MARVEL energy level accuracies, must be considered with care. Initial
uncertainties were estimated using a variant of Watson's relatively robust weighting \cite{04Watson} with final
results being based on bootstrap values.

MARVEL4 utilizes a bootstrap method to derive a final uncertainty for each empirical rovibrational energy levels \cite{jt908}. Each transition uncertainty is scaled by a random number between 1 and 10, and MARVEL4 is run for a fixed number of iterations. If the average of the bootstrapped energies differs by a statistically significant amount from the original MARVEL energy, or the standard deviation of the bootstrapped energies is larger than the original MARVEL energy uncertainty, MARVEL4 increases the uncertainty for this energy level. A more thorough discussion is given as part of the MARVEL investigation of N${_2}$O \cite{jt908},  which compares the bootstrap method to the previous method.

Previous MARVEL studies using the bootstrap method have suggested that the uncertainties generated are fairly insensitive to changes beyond 100 iterations \cite{jt908} - this has therefore been used for the CS${_2}$ project. We note that the MARVEL input format has two uncertainty columns:
the first column is the original or input uncertainty. MARVEL can change (increase) the uncertainty in which case this revised value is
entered in the second column; this option was not used here.

\subsection{Quantum numbers and selection rules}
Unlike CO$_2$, the rovibrational states of CS$_2$ are generally characterized by the standard linear-molecule, harmonic oscillator (HO) notation. Four quantum numbers are used to denote the vibrational states. $v_1$ and $v_3$ correspond to the molecule's symmetric and antisymmetric stretch respectively, and $v_2$ describes the degenerate bending mode. An angular momentum arises due to the bend occurring in two orthogonal planes with different phases as if the excited molecule is rotating about the molecular axis. The angular momentum quantum number $\ell$ takes non-negative values $v_2, v_2 -2, v_2 -4, \dots$ 1 or 0. The vibrational state of the molecule can be grouped as $(v_1, v_{2}^{\ell}, v_3)$; denoted below as $(v_1\ v_{2}\ {\ell}\ v_3)$.

The quantum number $J$ describes the molecule's rotation, with $J\geq \ell$. Additionally, there is a rotationless parity $p$ associated with each quantum state, holding values $p=0$ (“e” state) or $p=1$ (“f” state) \cite{Brown}. As $^{12}$C and $^{16}$S both have zero nuclear spin meaning that \CS\ has no hyperfine splittings, and that $J$ and $p$ are exact quantum numbers, in contrast to the approximate vibrational quantum numbers. They follow a set of strict selection rules, outlined as follows:\\
$\Delta J = 0,\pm1$\\
$\Delta J = 0, e\leftrightarrow f,$  but with  $J = 0\rightarrow0 $ not allowed\\
$\Delta J = \pm 1, e\leftrightarrow e, f\leftrightarrow f$.\\
States with $\ell = 0$ always have a rotationless parity e, and in principle, states with $\ell\geq 1$ can have both e and f parity. In the case of symmetric CS$_2$, however, the Pauli principle leads to a constraint on the rotational levels which means that half of them are missing: to be present, the sum 
$(J + \ell + v_3 + p)$ must be even.
In most papers considered here, the $p$ quantum number was not specified; the above convention was used to label and verify line assignments.

\section{Results}

\nbTR\ transition lines were extracted from \nbREF\ literature sources.
Table~\ref{t:marvelin} gives an extract of the MARVEL input file in 
format used by MARVEL3 \cite{jt750} and MARVEL4 \cite{jt908}. The full file is given in the supplementary material.

\begin{table}[h]
\caption{Extract from the \CS\ MARVEL input file. Transition wavenumbers ($\nu$) and uncertainties
(unc) are given in the units specified by the segment file. The full transitions file and the segment file are given as supplementary material.}
\label{t:marvelin}
%\resizebox{\columnwidth}{!}{%
\begin{tabular}{ccccccccrcccccrc}
 \hline
 $\nu$&  unc & unc & $v_1'$& $v_2'$& $\ell'$& $v_3'$& $p'$ &$J'$ &$v_1''$& $v_2''$& $\ell''$& $v_3''$& $p''$ &$J"$& tag\\
 \hline
2185.030036 &0.000010 & 0.000010 & 1 & 0 &  0 &1 & e &    1 &  0 &0 &0 &0 &e &2 &20KaGoKoMu.01\\
2184.584939 &0.000010 & 0.000010 & 1 & 0 &  0 &1 & e &    3 &  0 &0 &0 &0 &e &4 &20KaGoKoMu.02\\
2184.132785 &0.000010 & 0.000010 & 1 & 0 &  0 &1 & e &    5 &  0 &0 &0 &0 &e &6 &20KaGoKoMu.03\\
2183.673771 &0.000010 & 0.000010 & 1 & 0 &  0 &1 & e &    7 &  0 &0 &0 &0 &e &8 &20KaGoKoMu.04\\
2183.207862 &0.000010 & 0.000010 & 1 & 0 &  0 &1 & e &    9 &  0 &0 &0 &0 &e &10 & 20KaGoKoMu.05\\
2182.735023 &0.000010 & 0.000010 & 1 & 0 &  0 &1 & e &    11 &   0 &0 &0 &0 &e &12 & 20KaGoKoMu.06\\
2182.255285 &0.000010 & 0.000010 & 1 & 0 &  0 &1 & e &    13 &   0 &0 &0 &0 &e &14 & 20KaGoKoMu.07\\
2181.768607 &0.000010 & 0.000010 & 1 & 0 &  0 &1 & e &    15 &   0 &0 &0 &0 &e &16 & 20KaGoKoMu.08\\
2181.275034 &0.000010 & 0.000010 & 1 & 0 &  0 &1 & e &    17 &   0 &0 &0 &0 &e &18 & 20KaGoKoMu.09\\
2180.774543 &0.000010 & 0.000010 & 1 & 0 &  0 &1 & e &    19 &   0 &0 &0 &0 &e &20 & 20KaGoKoMu.10\\
2180.267154 &0.000010 & 0.000010 & 1 & 0 &  0 &1 & e &    21 &   0 &0 &0 &0 &e &22 & 20KaGoKoMu.11\\
2179.752889 &0.000010 & 0.000010 & 1 & 0 &  0 &1 & e &    23 &   0 &0 &0 &0 &e &24 & 20KaGoKoMu.12\\
\hline
\end{tabular}
\end{table}

Table \ref{table:1} lists the experimental sources used to construct the {CS}$_2$ rovibrational SN of this study. The data given includes, for each source, the wavenumber ranges of the validated transitions (in cm$^{-1}$), the number of actual versus validated (A/V) transitions, and selected uncertainty statistics (in cm$^{-1}$), where AOU is the average original uncertainty, AMR is the average MARVEL reproduction of the source’s lines, and MR the maximum reproduction in the source. We note that in the A/V statistic, the validated number refers solely to transitions in the main network and components in the sub networks are not counted. Of the \nbTR\ original considered, 53 had to be removed as this were not validated by the MARVEL procedure. The remaining data contained \nbUTR\ unique transitions. Notes on our use of these 
experimental sources
are provided in the Appendix.

\begin{table}[h]
\caption{\CS\ experimental sources used to construct the MARVEL spectroscopic network; also given are the range of transition
frequencies in each source, the  actual versus validated (A/V) number of transitions and
the average original uncertainty (AOU), average MARVEL reproduction (AMR) of the source’s lines, and maximum
reproduction (MR) in the source.}
\resizebox{\columnwidth}{!}{%
\begin{tabular}{lccccr}
 \hline
 Tag & Range in \cm & A/V & AOU & AMR & MR\\
 \hline
00BlWaBrDu \cite{00BlWaBrDu} &  $3083.12420-3168.28350$ & 760/749 & 2.166e-04 & 1.702e-04 & 0.786\\
01BlWaBrDu \cite{01BlWaBrDu} &  $3402.60150-4242.43020$ & 1122/1122 & 3.000e-04 & 1.952e-04 & 0.651\\
04HoAnPiAl \cite{04HoAnPiAl} &  $240.55342-1236.83693$ & 607/593 & 1.195e-04 & 1.471e-04 & 1.231\\
12VaTsLeHe \cite{12VaTsLeHe} &  $6097.02420-6466.26510$ & 198/198 & 1.036e-02 & 8.070e-03 & 0.779\\
20KaGoKoMu \cite{20KaGoKoMu} &  $2140.27362-2198.93869$ & 688/688 & 1.000e-05 & 9.085e-06 & 0.909\\
70SmOv \cite{70SmOv} &  $1512.74500-1549.85800$ & 175/166 & 1.524e-03 & 5.495e-03 & 3.606\\
71SmOv \cite{71SmOv} &  $378.87400-412.33200$ & 70/69 & 3.000e-03 & 1.089e-02 & 3.629\\
73Maki \cite{73Maki} &  $2286.36930-2977.04550$ & 314/312 & 2.000e-03 & 1.261e-03 & 0.631\\
74MaSa \cite{74MaSa} &  $2125.25720-2198.89730$ & 511/502 & 1.181e-03 & 2.296e-03 & 1.944\\
79JoKaAn \cite{79JoKaAn} &  $245.39500-275.65770$ & 97/97 & 2.000e-04 & 4.716e-04 & 2.358\\
80JoKa \cite{80JoKa} &  $368.71100-430.26260$ & 551/536 & 6.000e-04 & 1.415e-03 & 2.358\\
84BaBlWaCo \cite{84BaBlWaCo} &  $842.50650-894.92090$ & 462/433 & 1.005e-03 & 7.301e-04 & 0.727\\
85BlBaCaWa \cite{85BlBaCaWa} &  $842.74040-892.89450$ & 353/299 & 5.000e-04 & 1.965e-04 & 0.393\\
85LiJo \cite{85LiJo} &  $1492.22240-1561.74990$ & 289/261 & 2.000e-04 & 2.549e-04 & 1.275\\
86DaBlWaCo \cite{86DaBlWaCo} &  $857.04020-893.80860$ & 56/56 & 1.000e-03 & 6.194e-04 & 0.619\\
88WeScMa \cite{88WeScMa} &  $1500.60364-1549.03823$ & 7/7 & 1.239e-04 & 1.232e-04 & 0.995\\
92BlWaBlBr \cite{92BlWaBlBr} &  $4489.59360-4572.26230$ & 682/681 & 3.000e-04 & 1.656e-04 & 0.552\\
92WaMaBl \cite{92WaMaBl} &  $689.69200-744.37560$ & 246/245 & 4.000e-04 & 4.380e-04 & 1.095\\
96PlMaHoDe \cite{96PlMaHoDe} &  $6438.74600-6466.42300$ & 58/56 & 5.000e-03 & 5.294e-03 & 1.059\\
99BlWaBrDu \cite{99BlWaBrDu} &  $5139.23960-5337.35300$ & 309/309 & 4.000e-04 & 2.778e-04 & 0.695\\
99LiHu \cite{99LiHu} &  $3302.78700-11995.63900$ & 1220/1220 & 1.000e-01 & 4.882e-02 & 0.488\\
\hline
\end{tabular}% 
}
\label{table:1}
\end{table}

The main MARVEL SN contained 8599 transitions. A total of \nbEL\ energy levels were given by the main network. Energy levels which are characterized by many transitions are both
secure and well-determined. 
Table~\ref{t:marvelout} gives an extract of the energy file produced by the MARVEL procedure;
the full file is given in the supplementary material. The uncertainties in this file are given by running the bootstrap procedure 100 times during the final MARVEL. These uncertainties are generally slightly higher than those given initially by MARVEL and summarized in Table~\ref{table:1}; this is probably reflection of slightly optimistic estimation of uncertainties in some of the experimental studies. We note that the bootstrap procedure does not alter the uncertainties associated with levels determined by a single transition, suggesting that these uncertainties may be underestimated.

\begin{figure}[t!]
\centering 
\includegraphics[width=0.95\textwidth]{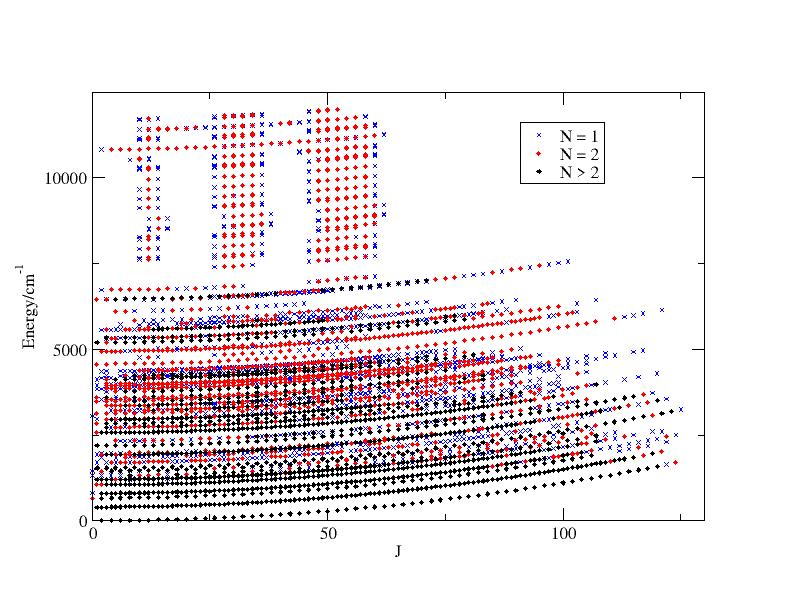}
\vspace{-1.25cm}
\caption{Summary of \CS\ energy levels  determined in this work designated by  $J$ and the 
number of transitions, $N$, determining them.}
\label{fig:levels}
\end{figure}

Figure~\ref{fig:levels} plots the energy levels as function of $J$ and energy; the plot distinguishes
between levels which are linked by only one transition ($N=1$), two ($N=2$) and many ($N>2$) transitions. 
In practice 1178 levels are characterized by a single transition, 1658 by two transitions and 1443 by more (in some cases
more than 40) transitions. While levels
characterized by only one transition cannot be regarded as secure, the generally smooth behavior of the levels as a function of $J$ suggests
that most of these levels are indeed correctly characterized.

\begin{table}[h]
\caption{Extract from the \CS\ energy levels output file giving standard quantum numbers, energies
with associated bootstrap uncertainty (100 iterations), plus the number of transitions ($N$) associated with each level. The full file is given as supplementary material.}
\label{t:marvelout}
%\resizebox{\columnwidth}{!}{%
\begin{tabular}{ccccrrrlr}
 \hline
 $v_1$& $v_2$& $\ell$& $v_3$& $p$ &$J$ &  $E$ / \cm & unc / \cm & $N$\\
 \hline
0 &0 &0 &0 &e &0 &    0.000000000  &   0.0        &   1233 \\
0 &0 &0 &0 &e &2 &    0.654706804  &   1.862e$-$05  &     25 \\
0 &0 &0 &0 &e &4 &    2.182361347  &   3.436e$-$05  &     29 \\
0 &0 &0 &0 &e &6 &    4.583103085  &   5.137e$-$05  &     36 \\
0 &0 &0 &0 &e &8 &    7.856854164  &   6.889e$-$05  &     42 \\
0 &0 &0 &0 &e &10&    12.003562632 &   9.015e$-$05  &     43 \\
0 &0 &0 &0 &e &12&    17.023231708 &   1.146e$-$04  &     44 \\
0 &0 &0 &0 &e &14&    22.915814696 &   1.388e$-$04  &     46 \\
0 &0 &0 &0 &e &16&    29.681277718 &   1.604e$-$04  &     44 \\
0 &0 &0 &0 &e &18&    37.319587303 &   1.792e$-$04  &     44 \\
0 &0 &0 &0 &e &20&    45.830710347 &   1.992e$-$04  &     46 \\
0 &0 &0 &0 &e &22&    55.214615363 &   2.186e$-$04  &     46 \\
\hline
\end{tabular}
\end{table}

\begin{table}[h]
\caption{Table of energy levels  with low $v_2$ and with  residues between MARVEL/Ames lists greater than 0.01\cm.}
\label{t:greater_0.01_residues}
\resizebox{\columnwidth}{!}{%
\begin{tabular}{llllccccrcc}
 \hline
 $N$ & $E$(Ames) / \cm & $E$(MARVEL) / \cm & unc(MARVEL) / \cm & $v_1$ & $v_2$ & $\ell$ & $v_3$ & $J$ & Obs(MARVEL) - Calc(Ames) / \cm \\
 \hline
2&1592.55085&1591.551616&0.001316&0&2&2&0&85&-0.99923\\
3&2477.47641&2477.457898&0.001439&0&2&0&1&37&-0.01851\\
1&3042.57405&3042.062948&0.003235&1&3&1&0&104&-0.51110\\
1&3088.72932&3088.715280&0.00327&1&3&1&0&106&-0.01404\\
1&3344.33572&3344.325423&0.002219&0&3&1&1&75&-0.01030\\
2&3632.40270&3632.324842&0.003949&1&3&1&1&50&-0.07786\\
1&3726.39141&3725.388616&0.002471&1&2&2&1&84&-1.00279\\
1&3766.90884&3766.919275&0.002147&1&3&1&1&61&0.01044\\
1&3794.08654&3794.166824&0.002155&1&3&1&1&63&0.08028\\
2&3880.94415&3880.933889&0.002195&1&3&1&1&69&-0.01026\\
1&3900.24745&3900.233803&0.002987&1&2&2&1&93&-0.01364\\
1&3998.73622&3998.99&0.1&6&0&0&0&28&0.25\\
1&4025.89849&4025.882898&0.002719&1&2&2&1&99&-0.01559\\
1&4163.10950&4163.021&0.1&3&6&2&0&14&-0.088\\
4&6554.57906&6554.565344&0.01365&3&0&0&3&29&-0.01372\\
3&6628.10881&6628.120680079&0.005015&3&0&0&3&39&0.01187\\
\hline
\end{tabular}}
\label{Table 5}
\end{table}

\begin{figure}[t!]
\centering 
\includegraphics[width=0.9\textwidth]{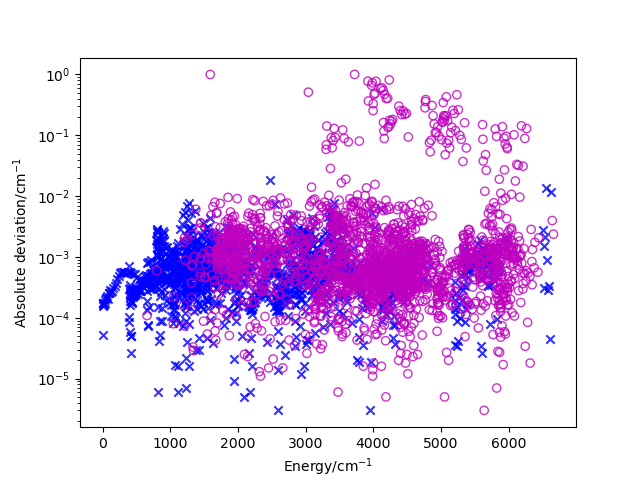}
\vspace{-0.5cm}
\caption{Plot of absolute residues between MARVEL/Ames lines against the associated MARVEL energy.
Blue crosses represent levels associated with 3 or more transitions; magenta circles represent levels with 2 or less transitions.}
\label{fig:abs_diff_energy}
\end{figure}

\begin{figure}[t!]
\centering 
\includegraphics[width=0.9\textwidth]{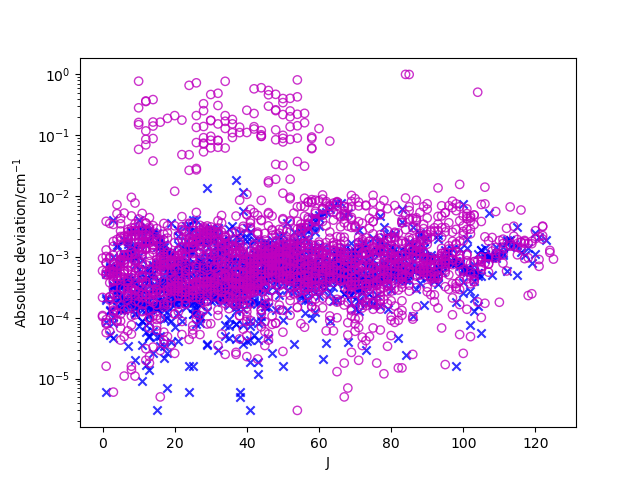}
\vspace{-0.5cm}
\caption{Absolute residues between MARVEL/Ames lines are plotted against the associated J value.
Blue crosses represent levels associated with 3 or more transitions; magenta circles represent levels with 2 or less transitions.}
\label{fig:abs_diff_J}
\end{figure}

\begin{figure}[t!]
\centering 
\includegraphics[width=0.9\textwidth]{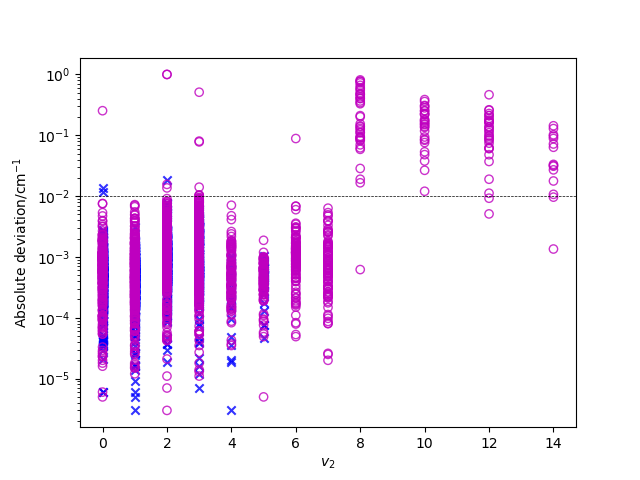}
\vspace{-0.5cm}
\caption{Plot of absolute differences against the associated $v_2$ quantum number. Blue crosses represent levels associated with 3 or more transitions; magenta circles represent levels with 2 or less transitions. Generally states with high values of $v_2$ seem to carry greatest differences.}
\label{fig:abs_diff_v2}
\end{figure}
\subsection{Comparison with NASA Ames line list}
The NASA Ames $^{12}$C$^{32}$S$_2$ line list \cite{24HuGoTa.CS2} contains 1~856~648 \CS\ rovibrational lines and covers the 0 - 6909 \cm wavenumber region. These transitions were generated using variational nuclear motion calculation with many of the energy levels replaced by the results of an Effective Hamiltonian model calculation \cite{22Tashkun}. 3494 of the \nbEL\ levels (66 of the 138 vibrational bands) underwent comparison; results of this comparison can be seen in Figs. 2, 3 and 4. Most of the vibrational bands not analyzed have band centers above the NASA Ames wavenumber range. However, three lower-lying bands, (0 4 0 3), (2 2 0 3) and (0 1 1 4), could not be verified by the Ames line list either. This is likely due to issues with quantum number labeling and  that these bands are not included in Tashkun's effective Hamiltonian model.
Almost all of the differences are smaller than 0.01 \cm; the deviations are larger than 0.01 \cm\ only for relatively high $v_2$ values. 

A comparison of the mean and standard deviations of the residues for low and high $v_2$ states is given i in Table ~\ref{t:s_dev_mean_tab}. There is 
a marked difference  between levels with $v_2 < 8$ and $v_2 \geq 8$. The standard deviation is bigger by one order of magnitude for high $v_2$ levels, and the mean is increased by approximately 2 orders of magnitude. 

On checking how many experimental measurements determine the MARVEL energy levels that have a large difference from the values, most of them are determined by only one or two measurements (see the magenta circle markers in Fig.~\ref{fig:abs_diff_energy} and Fig.~\ref{fig:abs_diff_v2}). 
While MARVEL levels determined by a single measurement are regarded as less trustworthy, the systematic correlation of high absolute deviations with high $v_2$ values suggests high $v_2$ states are not well represented in the Ames line list. 
A full list of levels which differ by more than
0.01 \cm\ between MARVEL and Ames is provided in the supplementary material; Table~\ref{t:greater_0.01_residues} provides a list of low $v_2$ levels
which differ by more than 0.01 \cm.\\

\begin{table}[h]
\caption{Comparison of MARVEL and Ames energy levels for low and high $\nu_2$ levels: standard deviation of signed deviations, and mean of signed and absolute deviations.}
\label{t:s_dev_mean_tab}
\begin{tabular}{lrr}
 \hline
 & $v_2 < 8$ & $v_2 \geq 8$ \\
 \hline
 Standard deviation (Obs(MARVEL)-Calc(Ames)) \cm & 0.0264 & 0.2413\\
 Mean (Obs(MARVEL)-Calc(Ames)) \cm & -0.0013 & -0.0927\\
 Mean (Absolute deviation) \cm & 0.0020 & 0.1844\\
\hline
\end{tabular}%
\label{Table 4}
\end{table}

Effective Hamiltonians can be an extremely effective and accurate method for representing experimental data. In favorable cases they can even improve on the uncertainty of individual measurements and can interpolate reliably between measured levels.
However, resonances, where levels from different vibrational bands with the same overall symmetry interact, can be 
problematic. MARVEL has no in principle difficulties with resonances but requires multiple transitions to a given level
for the energy to be securely determined unless there is an independent means of confirming them such as an independently computed line list. It is likely that at least some of the differences listed in Table~\ref{t:greater_0.01_residues} are caused
by resonances which have not been fully accounted for in the effective Hamiltonian treatment. 

\section{Conclusion}

We present a MARVEL analysis of $^{12}$C$^{32}$S$_2$ high resolution spectra which result in the determination of \nbEL\ empirical
energy levels. The uncertainties for these levels were determined by the implementation the new MARVEL4 bootstrap approach, with an iteration count of 100. Comparison of these levels with a recently produced NASA Ames line list \cite{24HuGoTa.CS2} generally gives very 
good agreement but identifies highly excited bending states as needing more work to better characterize them
by, for example, improving the representation  the higher bending states by an {\it ab initio} or fitted potential energy surface (PES). We note that MARVEL provides energy levels up to ones with $v_2=22$, so there are plenty of levels to test against
in constructing such a PES. The plan is to use the present
study as the starting point to generate a hot $^{12}$C$^{32}$S$_2$ line list as part of the ExoMol project; we note that the treatment of
excited bending states is particularly important for high temperatures as it is these states that give rise to many of the hot bands
that are key characteristic of hot line lists. While empirical energy levels provide important input to such a line list, experience
shows that transition intensities can be computed to high accuracy using {\it ab initio} dipole moment surfaces \cite{jt613,jt667}.

Further MARVEL studies could also be conducted on the some of the major  isotopologues of CS$_2$. Of particular interest for astronomy may be those molecules containing $^{13}$C, due to its high relative abundance in parts of the Universe. Although a fair number of experiments have been carried out into the rovibrational spectra of these isotopologues, it is unclear whether there is enough experimental data to form a robust MARVEL spectroscopic network.
\newpage

\section*{Supporting material}

The following are given as supporting material:\\
MARVEL input file - MARVEL4\_input\_CS2\_28Dec24.txt\\
A segment file necessary for running MARVEL4 - Marvel4\_segment\_CS2.txt.\\
MARVEL output energies file - MARVEL\_EnergyLevels\_28Dec24.txt\\
MARVEL AMES energy comparison file - Ames+MARVEL\_comparison\_28Dec24.txt\\
A file of levels differing by more than 0.01 \cm\ between MARVEL and the NASA Ames line list - Residues\_0.01\_28Dec.txt\\

\section*{Data availability statement}
All data is given in the article or the supporting materials.

\section*{Acknowledgement}
TS thanks UCL's Faculty of Maths and Physics Sciences (MAPS) for the provision of summer studentship.
This work was supported by ERC Advanced Investigator Project grant 883830.
\newpage

\bibliographystyle{elsarticle-num}
%\bibliography{journals_phys,jtj,CS2_MARVEL,CS2,MARVEL,OCS}

\appendix
\section{Notes on data sources}
\noindent\textbf{20KaGoKoMu} \cite{20KaGoKoMu}:The uncertainty is given in the paper as 'better than 0.00001 \cm'. This is the value used for all lines.\\\\
\textbf{12VaTsLeHe} \cite{12VaTsLeHe}: An uncertainty of 0.008~\cm\    was estimated from the standard deviation table for all lines. This was perhaps a little optimistic, as reflected in the initial MARVEL run. As a result, the uncertainties of the lines in the (0403) - (0000) band were increased to 0.026~\cm.
\\\\
\textbf{71SmOv} \cite{71SmOv}: The uncertainty was estimated to be 0.003~\cm\    for all lines. The R(60) line observed in the (0 1 1 0) - (0 0 0 0) transition was removed due to inconsistency.\\\\
\textbf{00BlWaBrDu} \cite{00BlWaBrDu}: An uncertainty of 0.002~\cm\    was assumed, estimated from the standard deviation table given. Two pairs of lines were removed due to self contradictory energy levels. The first pair consisted of the P(26) and R(24) lines in the (0 4 0 1) - (0 0 0 0) band, and the second pair was the R(86) and P(88) lines in the (0 5 1 1) - (0 1 1 0) band. \\\\
\textbf{96PlMaHoDe} \cite{96PlMaHoDe}: An experimental accuracy of 0.005~\cm\    was provided in the paper - this was taken to be the uncertainty. The P(18) and P(20) lines observed in the (3 0 0 3) - (0 0 0 0) transition were removed due to inconsistency.
\\\\
\textbf{73Maki} \cite{73Maki}: An uncertainty of 0.002~\cm\    was assumed, estimated from the standard deviation table. Some lines were labelled as 'Transitions to perturbed levels', however these uncertainties were kept the same. The R(47) line observed in the (1 3 1 1) - (0 1 1 0) transition was removed due to inconsistency.\\\\
\textbf{79JoKaAn} \cite{79JoKaAn}: An uncertainty of 0.0002~\cm\    was assumed, estimated from the standard deviation given for the band centres.\\\\
\textbf{84BaBlWaCo} \cite{84BaBlWaCo}: The uncertainty was taken to be 0.001 cm$^{-1}$, as mentioned in the paper. The P(97) line observed in the (0 1 1 1) - (1 1 1 0) band was removed due to inconsistency.
The uncertainties for two lines in the (0 2 0 1) - (1 2 0 0) transition band were doubled - these lines were both assigned J' = 37.\\\\
\textbf{85BlBaCaWa} \cite{85BlBaCaWa}: An uncertainty of 0.0005~\cm\    was assumed, estimated from the standard deviation table given.\\\\
\textbf{85LiJo} \cite{85LiJo}: An uncertainty of 0.002~\cm\    was assumed, estimated from the standard deviation table. The R(64) line observed in the (1 0 0 1) - (1 0 0 0) transition was removed due to inconsistency.\\\\
\textbf{92BlWaBlBr} \cite{92BlWaBlBr}: An uncertainty of 0.0003~\cm\    was assumed, estimated from the standard deviation table given.\\\\
\textbf{92WaMaBl} \cite{92WaMaBl}: An uncertainty of 0.0004~\cm\    was assumed, estimated from the standard deviation table given.  The R(95) line observed in the (0 1 1 1) - (0 3 1 0) transition was removed due to inconsistency.
\\\\
\textbf{99BlWaBrDu} \cite{99BlWaBrDu}: An uncertainty of 0.0004~\cm\    was assumed, estimated from the standard deviation table given.
\\\\
\textbf{01BlWaBrDu} \cite{01BlWaBrDu}: An uncertainty of 0.0003~\cm\    was assumed, estimated from the standard deviation table given. The R(64) line in the (1 5 1 1) - (0 1 1 0) transition was removed due to inconsistencies.
\\\\
\textbf{04HoAnPiAl} \cite{04HoAnPiAl}: The uncertainties, as assigned to every line, were adopted directly from the paper. These proved to be very optimistic, so every uncertainty was scaled up by a factor of 10. Four particularly bad lines plus 15 lines belonging to the (0 3 1 0) - (0 0 0 0) and (0 1 1 0) - (0 0 0 0) bands were removed due to inconsistencies.\\\\
\textbf{70SmOv} \cite{70SmOv}: An uncertainty of 0.001~\cm\    was assumed, estimated from the standard deviation table. 9 lines belonging to the (0 0 0 1) - (0 0 0 0) band were removed due to inconsistencies.\\\\
\textbf{74MaSa} \cite{74MaSa}: An uncertainty of 0.001~\cm\    was estimated from the standard deviation table. This was doubled for transitions labelled in the paper as 'weak/blended'. The P(70) and R(68) lines observed in the (2 1 1 1) - (1 1 1 0) transition were removed due to inconsistency. The R(50) line observed in the (1 2 2 1) - (0 2 2 0) transition was also removed.\\\\
\textbf{80JoKa} \cite{80JoKa}: The uncertainty was estimated to be 0.0006~cm$^{-1}$, from the standard deviation table. This was a little optimistic, as reflected in the MARVEL bad line analysis. The MARVEL4 bootstrap method was used to calculate improved uncertainties for the associated energy levels. 15 lines belonging to the (0 3 1 0) - (0 2 2 0), (0 3 1 0) - (0 2 0 0), (1 1 1 0) - (1 0 0 0) and (0 2 2 0) - (0 1 1 0) bands were removed due to inconsistencies while two lines were corrected for obvious
typographic errors.
\\\\
\textbf{88WeScMa} \cite{88WeScMa}: This paper contains measured transitions in MHz - the MARVEL4 segment file was updated to reflect this. The uncertainties were obtained directly from the paper for each line.
\\\\
\textbf{86DaBlWaCo} \cite{86DaBlWaCo}: This paper contained only spectral intensities, and raw transition data was not publicly available. However, the line positions are given in Tashkun's recent paper using  an effective Hamiltonian to construct a line list \cite{22Tashkun}  and the attached data file was used as a secondary source.\\\\
\textbf{99LiHu} \cite{99LiHu}: This paper reports stimulated emission spectra which provide observed energy levels for both the ground and R~$^3$B$_2$ excited states. Although the excited state vibrational bands were ambiguously assigned, there are clear data for the ground levels. The lower level of the transition was taken to be ($\nu_1, \nu_2, l, \nu_3, J$) = (0 0 0 0 0) in each case.
The uncertainty was set to 0.1~\cm\    for every line. This was taken from the data file attached to the recent paper by  Tashkun \cite{22Tashkun}, which used an effective Hamiltonian to model line positions for {CS}$_2$.
\end{document}